\documentclass[proof]{pasj00}
\draft
      
\SetRunningHead{Haze et al.}{Multi-Color Coronagraph Experiment in a Vacuum Testbed with a Binary Shaped Pupil Mask}
\Received{2010 October 23}
\Accepted{2011 May 3}
\Published{出版日}

\begin{document}

\title{Multi-Color Coronagraph Experiment in a Vacuum Testbed with a Binary Shaped Pupil Mask}


\author{
   Kanae \textsc{Haze}\altaffilmark{1,2}
   Keigo \textsc{Enya}\altaffilmark{2}
   Lyu \textsc{Abe}\altaffilmark{3}
   Takayuki \textsc{Kotani}\altaffilmark{2}
   Takao \textsc{Nakagawa}\altaffilmark{2}
   Toshimichi \textsc{Sato}\altaffilmark{4}
   and
   Tomoyasu \textsc{Yamamuro}\altaffilmark{5}}
 \altaffiltext{1}{JSPS Research Fellow, Department of Space and Astronautical Science,
        The Graduate University for Advanced Studies, 3-1-1 Yoshinodai, Chuo-ku, Sagamihara 252-5210}
 \email{haze@ir.isas.jaxa.jp}
 \altaffiltext{2}{Department of Infrared Astrophysics, 
        Institute of Space and Astronautical Science, Japan Aerospace Exploration Agency (JAXA), 
        3-1-1 Yoshinodai, Chuo-ku, Sagamihara 252-5210}
 \altaffiltext{3}{Laboratoire Hippolyte Fizeau,
        UMR 6525 Universit\'e de Nice-Sophia Antipolis,
        Parc Valrose, F-06108 Nice, France}
 \altaffiltext{4}{Nanotechnology Research Institute,
       Advanced Industrial Science and Technology, 1-1-1 Azuma, Tsukuba, Ibaraki 305-8561}
 \altaffiltext{5}{Optcraft, 3-26-8 Aihara, Sagamihara, Kanagawa 229-1101}


%
 

\maketitle

\begin{abstract}

We conducted a number of multi-color/broadband coronagraph 
experiments using a vacuum chamber and a binary-shaped pupil mask 
which in principle should work at all wavelengths, 
in the context of the research and development on a stellar coronagraph 
to observe extra-solar planets (exoplanets) directly. 
The aim of this work is to demonstrate that 
subtraction of Point Spread Function (PSF) and 
multi-color/broadband experiments using a binary-shaped 
pupil mask coronagraph would help improve the contrast 
in the observation of exoplanets. 
A checker-board mask, a kind of binary-shaped pupil mask, was used. 
We improved the temperature stability by installing 
the coronagraph optics in a vacuum chamber, 
controlling the temperature of the optical bench, and covering the 
vacuum chamber with thermal insulation layers. 
Active wavefront control is not applied in this work. 
We evaluated how much the PSF subtraction contributes to 
the high contrast observation by subtracting the images obtained 
through the coronagraph. 
We also carried out multi-color/broadband experiments 
in order to demonstrate a more realistic observation 
using Super luminescent Light Emitting Diodes (SLEDs) with center 
wavelengths of 650nm, 750nm, 800nm and 850nm. 
A contrast of 2.3×$10^{-7}$ was obtained for the raw 
coronagraphic image and a contrast of 1.3$\times 10^{-9}$ 
was achieved after PSF subtraction with a He-Ne laser at 632.8nm 
wavelength. 
Thus, the contrast was improved by around two orders of magnitude from the raw contrast 
by subtracting the PSF. 
We achieved contrasts of 
3.1$\times 10^{-7}$, 1.1$\times 10^{-6}$, 1.6$\times 10^{-6}$ and 2.5$\times 10^{-6}$ 
at the bands of 650nm, 750nm, 800nm and 850nm, respectively, in the multi-color/broadband experiments. 
The results show that contrast within each of the wavelength bands was significantly improved compared with non-coronagraphic optics. 
We demonstrated PSF subtraction is potentially beneficial for improving contrast of a binary-shaped pupil mask 
coronagraph, 
and this coronagraph produces a significant improvement in contrast 
with multi-color/broadband light sources.

\end{abstract}

\section{Introduction}

%
   Direct detection and spectroscopy of extra-solar planets (exoplanets) is essential for understanding 
   how planetary systems were born, how they evolve, and, ultimately, 
   for finding biological signatures on these planets. 
   The enormous contrast in luminosity between the central star and a planet presents the primary difficulty 
   in the direct observation of exoplanets. 
   For example, if the solar system is observed from a distance, 
   the expected contrast between the central star and the planet at visible light wavelengths 
   is $\sim10^{-10}$ but is reduced to $\sim10^{-6}$ in the mid-infrared region \citep{Traub}. 
   In this case, it is difficult to make direct observations with an ordinary telescope 
   because the planet is buried in the halo of the central star image. 
   Recently, some studies on direct observations were presented \citep{Marois2008, Kalas2008}, 
   but only if they are young gas-giant planets very far from the central star. 
   We started, therefore, to study the stellar coronagraph, first reported by \cite{Lyot}, 
   which is capable of very high contrast observations. 
   With this we aim to systematically study the exoplanets. 
   Of the various kinds of coronagraphs, we focused on a binary-shaped pupil mask coronagraph. 
   The reasons for using this coronagraph are 
   that it is robust against pointing errors, 
   is essentially achromatic and 
   is relatively simple 
   \citep{Jacquinot, Spergel, Vanderbei2003, Vanderbei2004, Green, Tanaka, Kasdin}. 
   The adoption of a binary-shaped pupil mask coronagraph 
   for the Space Infrared telescope for Cosmology and Astrophysics (SPICA) coronagraph 
   are considered \citep{Enya2008b, Nakagawa}. 

  We carried out experiments to investigate the performance of the coronagraph using a binary-shaped pupil mask 
  in the visible light region \citep{Enya2007a, Enya2008a}. 
  The coronagraph optics were installed in a vacuum chamber to reduce the instabilities 
  caused by air turbulence and thermal instability. 
  We also demonstrated the use of the point spread function (PSF) subtraction method to improve contrast \citep{Haze}. 
  A He-Ne laser was employed as the light source in the experiments. 
  In principle, the binary-shaped pupil mask coronagraph should work at all wavelengths. 
  For an actual observation, it is necessary to make observations over a wavelength band and 
  it would be beneficial to make observations using multiple bands. 
  PSF subtraction can be applied between different wavelengths in multi-band imaging. 
  This method has the advantage of improving the dynamic range by removing the starlight and speckle patterns \citep{Biller}. 
  In this paper, we present the improved results of a laboratory experiment using a binary-shaped pupil mask coronagraph 
  installed in a vacuum chamber, the result of using PSF subtraction, 
  and our first results of a demonstration using multi-color/broadband light sources. 
  
%
%
%
%
%

\newpage

\section{Experiment}
%
%
%
%
  %
  %
  %
  In this section, we present two kinds of experiment, subtraction of PSF and multi-color/broadband demonstrates. 
  We describe the common part of the two experiments and then describe each separately. 

\begin{figure*}[tbp]
                                      \begin{center}
                                       \FigureFile(80mm,80mm){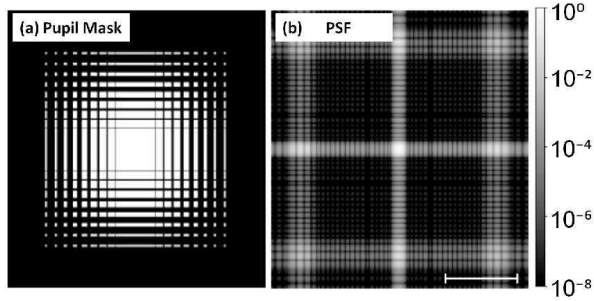}
                                      \end{center}
                                      \caption{Panel (a): the pupil mask design. 
                                      The transmissivity through the black and white regions is 0 and 1, respectively. 
                                      The diameter of the circumscribed circle to the transmissive part is 2 mm. 
                                      Panel (b): the expected (theoretical) PSF by the pupil mask. 
                                      The scale bar corresponds to 20 $\lambda /D$.
                                      }
                                      \label{fig1}
                                     \end{figure*}

      We use Mask2 developed in \cite{Enya2007a}. 
      The mask is a checkerboard mask, which is a type of binary-shaped pupil mask \citep{Vanderbei2004}. 
      Fig.\ref{fig1} shows the design. 
      The central brightest region of the PSF is called the "core", 
      and the four regions near to the core, in which diffracted light is reduced, is called the "dark region". 
      The contrast, inner working angle (IWA), and outer working angle (OWA) are 
      $10^{-7}$ , $3\lambda/D$, and $30\lambda/D$, respectively, 
      where $\lambda$ is wavelength and 
      $D$ is the diagonal of the checkerboard mask. 
      Optimization of the mask shape was performed using the LOQO solver presented by \cite{Vanderbei1999}. 
      The pupil mask consists of a 100nm thick aluminum film on a BK7 substrate, 
      and was manufactured using nano-fabrication technology 
      at the National Institute of Advanced Industrial Science and Technology (AIST) in Japan. 
      A standard broadband multiple-layer anti-reflection (BMAR) coating optimised for use at a wavelength of 632.8nm 
      was applied to both sides of the substrate. 

\begin{figure*}[tbp]
                                             \begin{center}
                                              \FigureFile(160mm,160mm){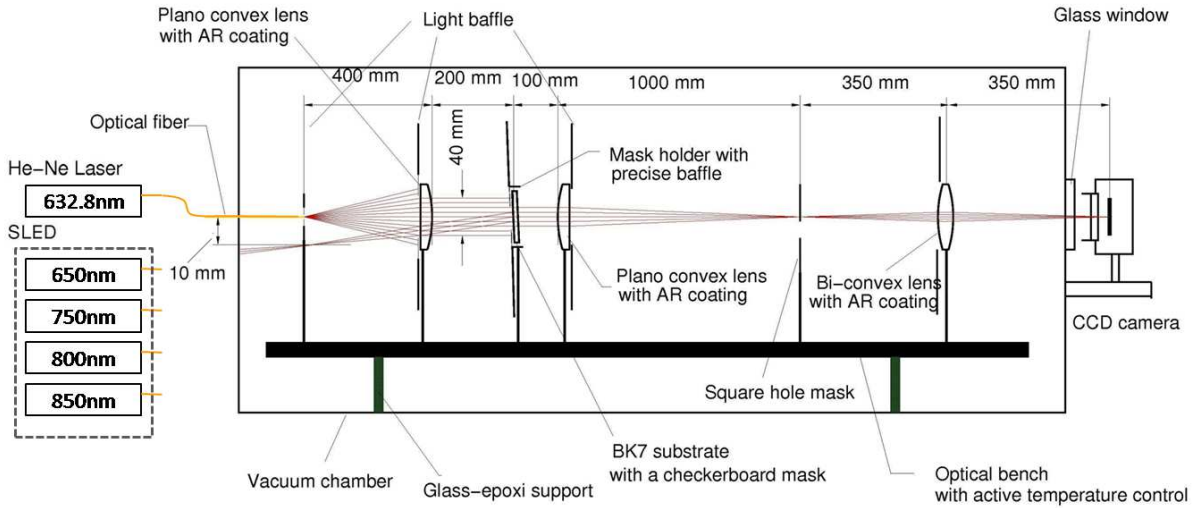}
                                             \end{center}
                                             \caption{The configuration of the experimental optics. }
                                             \label{fig2}
                                            \end{figure*}

        Fig.\ref{fig2} shows the instrument used for this work. 
        All the experimental optics were located in a clean-room 
        at the Institute of Space and Astronautical Science/ Japan Aerospace Exploration Agency (ISAS/JAXA). 
        The coronagraph optics were set up in a vacuum chamber. 
        We used a He-Ne laser with a wavelength of 632.8nm and Super luminescent Light Emitting Diodes (SLEDs) 
        with center wavelengths of 650nm, 750nm, 800nm and 850nm, as light sources (Table \ref{table1}). 
        Light is passed into the chamber through a single-mode optical fiber. 
        The entrance beam from the optical fiber is collimated by a 50mm diameter BK7 plano-convex lens, 
        and the collimated beam passing through the pupil mask is focused by a second plano-convex lens. 
        Active wavefront control is not applied in this work. 
        A bi-convex lens is used to reproduce the image through a window in the chamber.
        BMAR coating optimized for wavelengths of 400-700 $\mu$m 
        has been applied to both sides of the lens and the window to reduce reflection at the surface. 
        A commercially available cooled CCD camera (BJ-42L, BITRAN) 
        with $2048\times2048$ pixels installed outside of the chamber is used to measure the PSF. 

        To obtain a high-contrast image, we carried out the following procedure.
        We measured the core and the dark region, 
        each of which have different imaging times, separately. 
        When the dark region was measured, we obscured the light from the core 
        with a square hole mask inserted at the first focal plane after the pupil mask. 
        When the core was measured, we removed the square hole mask and put in two neutral density (ND) filters. 
        The transmission through the ND filters is dependent on wavelength. 
        We show the measurements of transmission at each wavelength in Table \ref{table1}. 

        \subsection{PSF subtraction experiment}
        Subtraction of PSF is beneficial in that it removes the static wavefront error (WFE) 
        and achieves a higher contrast than the raw contrast of the coronagraph \citep{Trauger}. 
        In space telescopes, the WFE caused by imperfections in the optics is 
        an important limiting factor in the contrast of a coronagraph. 
        PSF subtraction is available in the direct observation of exoplanets using space telescopes 
        and helps improve the high contrast observation. 
        We demonstrated how much the PSF subtraction contributes to the high contrast observation with the checkerboard mask
        in the laboratory under vary stable environment. 
        The variations of WFE in the laboratory experiment is attributed to thermal deformation of the optics. 
        To maximize the effects of the PSF subtraction, it is important to improve the thermal stability of 
        the entire coronagraph optics. 
        
        %
        %
        %
        To reduce the instabilities caused by air turbulence and thermal instability, 
        the coronagraph optics were installed on an optical bench set in a vacuum chamber. 
        The optical bench was set on a support consisting of glass epoxy plates 
        to prevent thermal conduction from the chamber.
        To reduce the thermal instability, active PID temperature control was applied to the optical bench using 
        a silicon diode temperature sensor and three resistance heaters. 
        These are described in our previous study \citep{Haze}. 
        In this study, we added new functions to the experimental system as follows. 
        We used multi-layer insulation (MLI), consisting of an aluminum film on a polyethylene foam mattress, 
        on the windows in the experiment room to reduce radiation from the windows. 
        In addition, we controlled the room temperature with an air conditioner and 
        installed fans to obtain thermal equilibrium through convection. 
        To reduce thermal deformation, 
        the whole external surface of the vacuum chamber was covered with eight sheets of MLI. 
        We monitored the change of temperature on the surface of the vacuum chamber and the camera stage 
        using a total of eight silicon diode temperature sensors (DT-670A1-SD, LakeShore) 
        and a temperature monitor (Model 218, LakeShore). 
        Before introducing the MLI, a cyclical change in temperature with a period of about 1000sec was observed in all eight channels 
        and the temperature variation range was about 0.1K (peak-to-peak). 
        After introducing the MLI, this cyclical change in temperature was not observed in any of the channels and 
        the temperature variation range in every channel was significantly reduced to less than 0.05K (peak-to-peak). 
        Consequently, we were able to ascertain whether the thermal stability had an effect on the PSF subtraction 
        in our experiments. 
        
\begin{figure}[tbp]
                                             \begin{center}
                                              \FigureFile(80mm,80mm){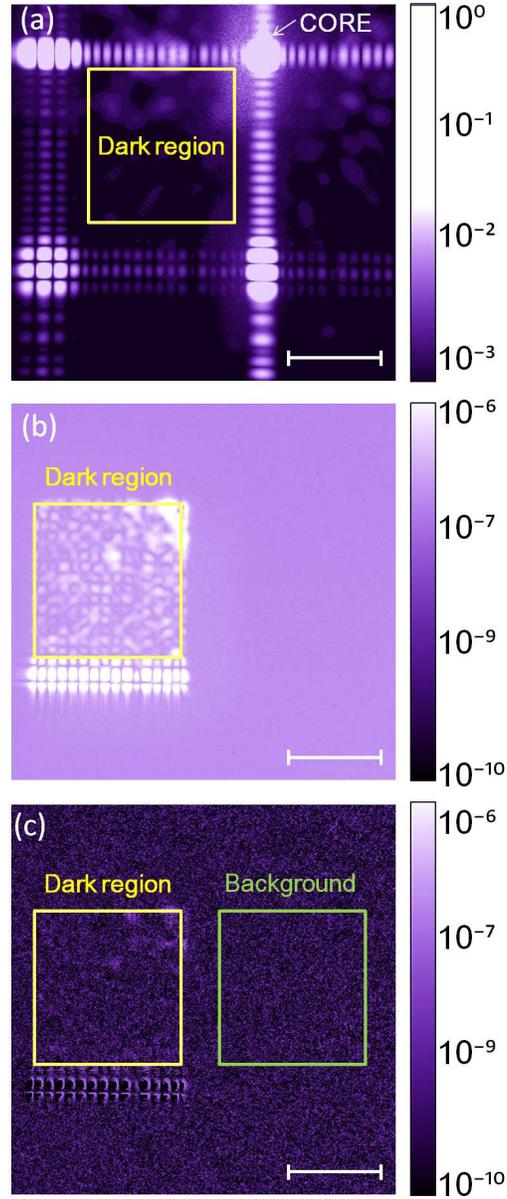}
                                             \end{center}
                                           　  \caption{Observed coronagraphic images obtained with He-Ne laser. 
                                               Panel (a): an image including the core of the PSF obtained with the ND filters. 
                                               Panel (b): a raw image of the dark region (the yellow rectangle) obtained with the square hole mask. 
                                               Panel (c): an image of the result of PSF subtraction using two raw coronagraphic images. 
                                               The green rectangle shows the area used to derive the background level in analysis. 
                                               The scale bars correspond to $10\lambda/D$.}
                                             \label{fig4}
                                            \end{figure}

%
We used a He-Ne laser in this experiment. 
The core images of the coronagraphic PSF were taken with a combination of several exposure times (0.03, 0.3, 3, 10s). 
The CCD was cooled and stabilized at $271.0\pm 0.5K (1\sigma)$. 
We inserted two ND filters as previously mentioned. 
After each imaging process, the laser source was turned off and a "dark frame" 
measurement was taken with the same exposure time and the same optical density filter. 
The dark frame was subtracted from the image with the laser light on and we obtained a "raw" coronagraphic image (Fig.\ref{fig4}a). 
This result is quite consistent with that expected from theory, as shown in Fig.\ref{fig1}b. 

%
%
The dark region of the coronagraphic image was observed with a 200s exposure.
A "dark frame" was taken with a 200s exposure 
and this was then subtracted from the dark region image with the laser light on. 
The CCD was at the same temperature as when the core images were observed. 
The observed dark region of the raw coronagraphic image is shown in Fig.\ref{fig4}b. 
The "Lattice pattern" consisting of dots in the "raw" dark region corresponds to the diffraction pattern caused by 
the checkerboard mask and hence it is limited by the design of the mask. 
In other words, the current experimental contrast approximately reached the design value. 
We evaluated the contrast between the areal mean of the observed dark region and the peak of the core. 
The observed dark region is the area of the image through the square hole mask. 
As a result, a raw contrast of 2.3×$10^{-7}$ was obtained. 

\begin{figure}[tbp]
                       \begin{center}
                        \FigureFile(80mm,80mm){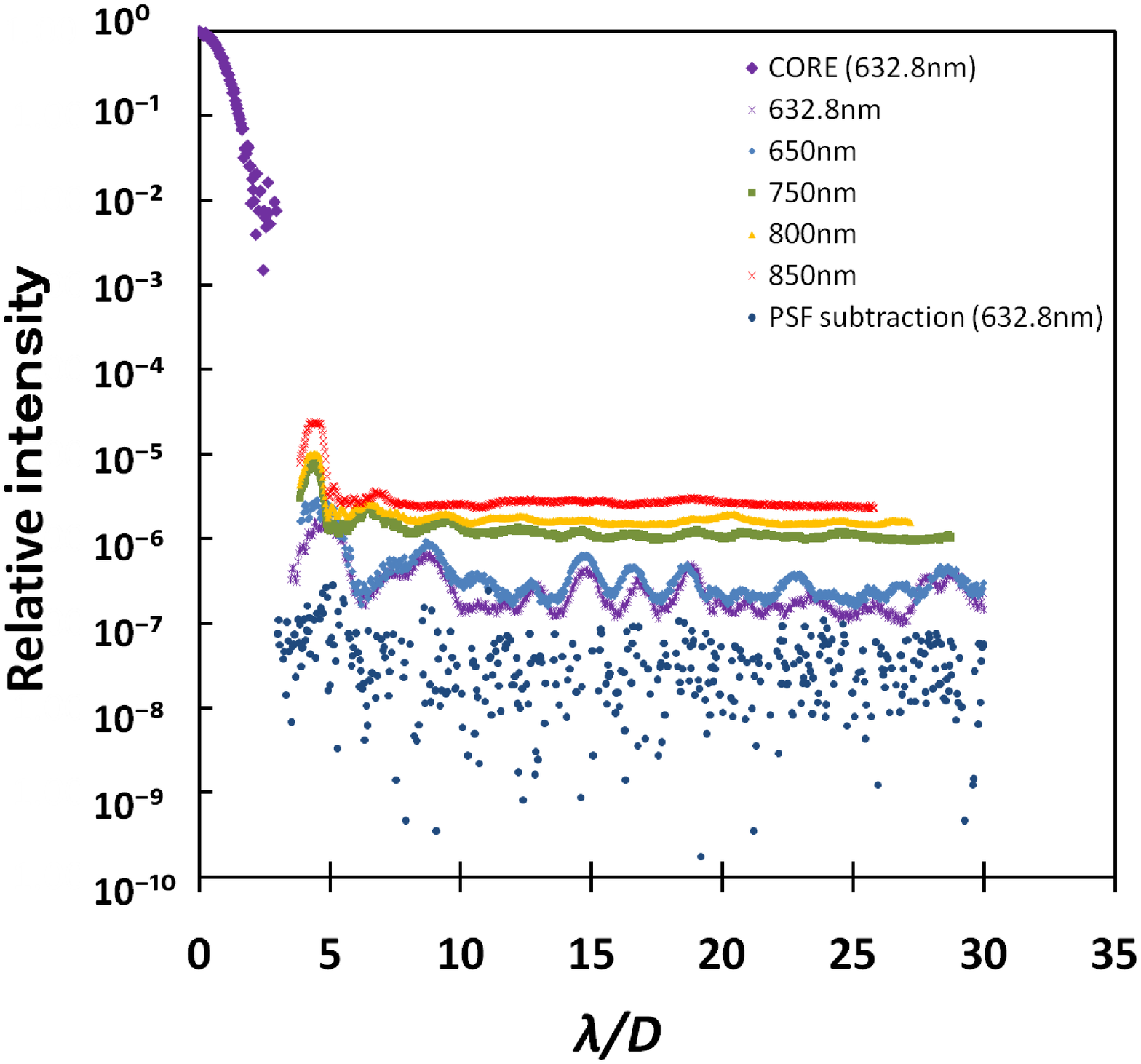}
                       \end{center}
                       \caption{Diagonal profiles of the observed coronagraphic PSF. 
                       Each profile is normalized by the peak intensity. 
                       }
                       \label{fig5}
                      \end{figure}

To obtain a contrast better than the design limit, we introduced PSF subtraction. 
The dark region of the coronagraphic image was observed 
in sets of $200s \times 18$ exposures (3600s) 2 times.
18 frames in each set of dark regions were combined and as a result 
two images totaling 3600s exposure were obtained. 
We obtained the dark region of the result of PSF subtraction using these two images. 
The images after PSF subtraction are shown in Fig.\ref{fig4}c.
The dark region consists of a "lattice pattern" derived from the design of the mask, 
"speckle" from systematic errors in the experiment 
and "random noise" including dark current and readout noise. 
The background consists of random noise.
Therefore, the standard deviation (1$\sigma$) of the coronagraphic image after PSF subtraction is given by 
$\sigma_{}=\sqrt{\sigma_{DR}^{2}-\sigma_{BG}^{2}}$, 
where $\sigma_{DR}$ is the standard deviation of the dark region 
and $\sigma_{BG}$ is the standard deviation of the background. 
We evaluated the contrast between the $\sigma$ after PSF subtraction and the peak of the core. 
A contrast of 1.3$\times 10^{-9}$ was achieved for the PSF subtracted image. 
Fig.\ref{fig5} shows one dimensional profiles of the coronagraphic images obtained by measurement. 
Scaling by exposure time and optical density allowed smooth profiles of the core to be obtained. 

 We considered detection threshold in order to carry out a practical comparison of the contrast 
 before and after the PSF subtraction. After the PSF subtraction, 
 the contrast was estimated using the standard deviation of the dark region, 
 as adopted in \citet{Biller}.
In our data, Gaussian distribution was confirmed by plotting values of the dark region 
and 1$\sigma$ cover $\sim70\%$ of values of the dark region. Before the PSF subtraction, 
the contrast was estimated using the areal mean of the dark region. 
We checked the values of the observed dark region and confirmed 
that the areal mean also covers $\sim70\%$ of values of the dark region. 
Therefore, we estimate that the contrast was improved 
by almost two orders of magnitude when compared with the raw PSF.
In other words, it indicates that the adiabatic vacuum chamber was successful in reducing the instabilities 
caused by thermal deformation of the optics reported previously \cite{Haze}. 

%
%
%
%
%
%
%
        \subsection{Multi-color/broadband experiment with SLED}
              For an actual observation, it is necessary to observe over a wavelength band and 
              it is beneficial to observe  multiple bands. 
              In principle, the binary-shaped pupil mask coronagraph should work at all wavelengths. 
              To demonstrate this, the light source was changed from a He-Ne laser 
              to broadband and multi-band light sources. 
              SLEDs from EXALOS 
              were used as multi-band and broadband light sources at four wavelengths. 
              The wavelengths and bandwidths are presented in Table \ref{table1}. 
              The CCD was cooled and stabilized at $271.0\pm 0.5K (1\sigma)$. 
              The temperature of CCD could not be reduced to the temperature used for the PSF subtraction experiment 
              and we assume that was because of a functional problem in the CCD cooler. 
              The wavelength was changed by connecting the optical fiber outside the chamber to each light source in turn. 
              For each of the four wavelengths, the core and dark region images were observed 
              without reconnecting the light source in order to prevent changes in intensity. 
              
\begin{figure*}[tbp]
                                             \begin{center}
                                              \FigureFile(130mm,130mm){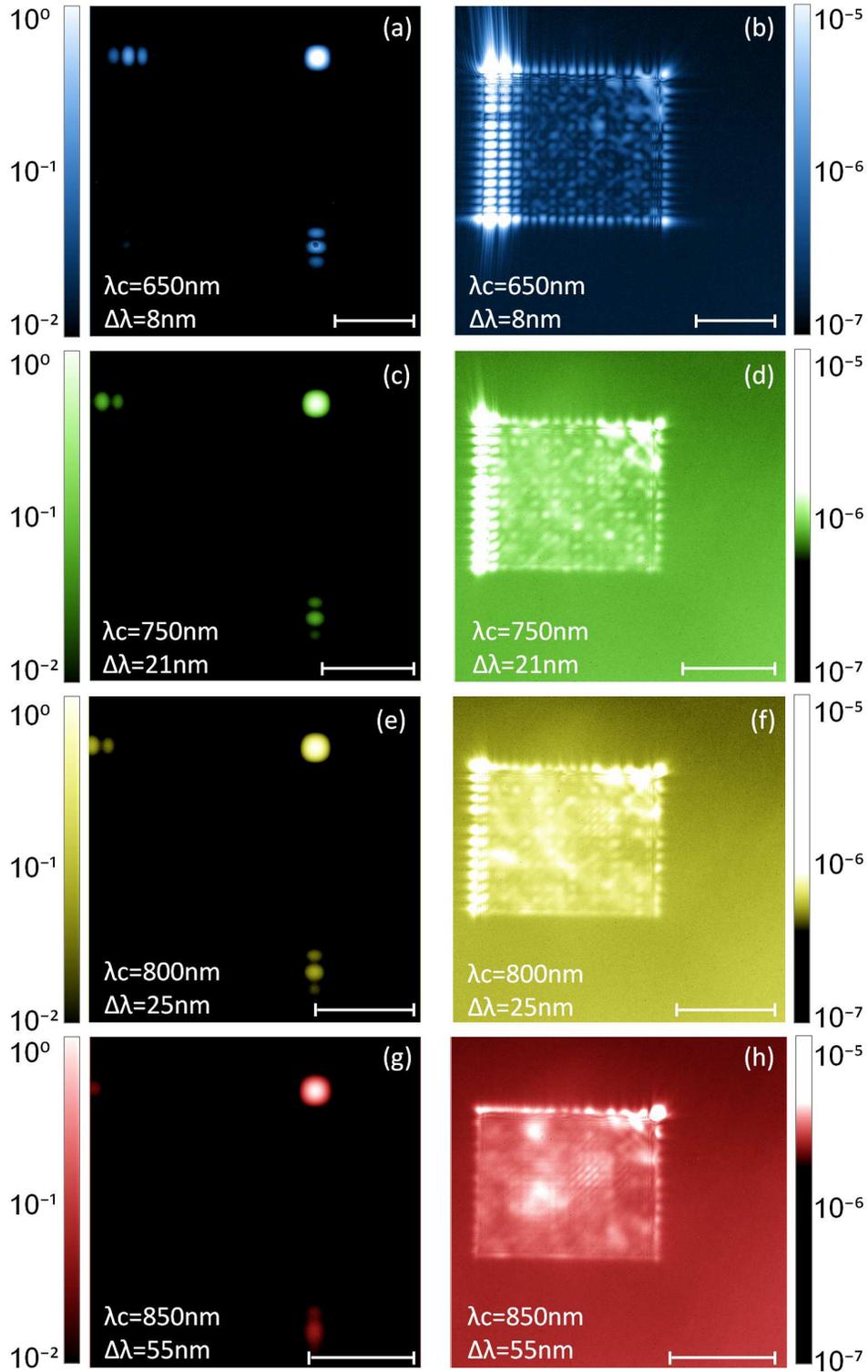}
                                             \end{center}
                                             \caption{Observed coronagraphic images obtained with SLEDs. 
                                             Panels (a), (c), (e), and (g) show images including the core of the PSF obtained with the ND filters, 
                                             and panels (b), (d), (f) and (h) show images of the dark region obtained with the square hole mask, 
                                             respectively. 
                                             The scale bars corresponds to  $10\lambda/D$. 
                                             }
                                             \label{fig3}
                                            \end{figure*}

              The resulting images are shown in Fig.\ref{fig3}. 
              The core images were taken with several exposure times (0.3, 0.6, 0.9, 1.2, 12, 24s) and 
              the dark region images were taken with 120s and 300s exposures. 
              As  mentioned previously, we inserted two ND filters when we took the core images.
              A "dark frame" was then subtracted from the image with the SLED light on. 
              As a result, 
              we achieved contrasts of 
              3.1$\times 10^{-7}$, 1.1$\times 10^{-6}$, 1.6$\times 10^{-6}$ and 2.5$\times 10^{-6}$ at 650nm, 750nm, 800nm and 850nm, respectively (Fig.\ref{fig5}). 
              The results show that the contrast was significantly improving compared with non-coronagraphic optics within each of the wavelength bands. 
              However, the contrast degrades as the wavelength gets longer. 
              We can see ghost images at longer wavelengths (Figs.\ref{fig3} b,d,f,h). 
              
 %
    	       %
    	       %
    %
    %
    %
    %
%

\section{Discussion}
         As a result of using PSF subtraction, we improved the contrast by around two orders of magnitude from the raw contrast 
         at the He-Ne laser wavelength. 
         The result confirmed that 
         PSF subtraction of stable images is potentially beneficial for improving contrast in the laboratory. 
         It should be noted that the result shows stability of our apparatus. 
         The stability and understanding of it are expected to be valuable 
         in more complex future experimental environment with the chamber. 
         If enough stability and improvement in contrast 
         thanks to PSF subtraction can be expected during coronagraph observations with 
         SPICA or other telescopes, 
         the requirements for the raw contrast can be significantly relaxed. 
         Thus, the requirements for the static WFE can also be relaxed in a stable environment. 
         This therefore can provide a significant advantage for the development of telescopes. 
         
 %
         %
         %
         %
           The first results of our demonstration using multi-color/broadband sources have shown that 
           the binary-shaped pupil mask coronagraph produces significant improvement in contrast at various wavelength bands, 
           compared with non-coronagraphic optics. 
           
          %
          %
%
         We also found the contrast degrades as the wavelength gets longer and a ghost image was observed at longer wavelengths. 
         Our suggestions on these issues are as follows.

　　　　　　　　　　　\begin{figure}[tbp]
                       \begin{center}
                        \FigureFile(80mm,80mm){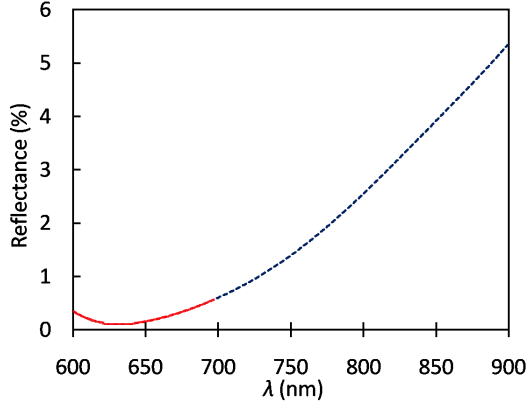}
                       \end{center}
                       \caption{Residual reflectance of BMAR coating. 
                       Red solid and blue dash lines are plots based on experiment and theoretical value, respectively.
                       }
                       \label{fig6}
                      \end{figure}

\begin{enumerate}
\item Wavelength dependency of the focal position : 
In this experiment, we changed each wavelength by reconnecting the optical fiber outside the chamber 
without changing the configuration of the optical system inside the chamber, which had been adjusted for the He-Ne laser. 
The focal planes at the various wavelengths did not coincide 
because the lenses have different refractive indices for different wavelengths of light. 
Therefore, we checked whether a major cause of the contrast deterioration 
was due to reducing the peak of the core by defocusing. 
As a result of simple ray trace analysis with ZMAX software, 
the difference of the focal plane position in the entire optical system was up to 60.6mm 
and the contrast at 850nm is deteriorated from the contrast at 632.8nm by factor of only 1.32.  
It was not a major cause of the contrast degradation.\\
\item Wavelength dependency of the residual reflectance of the BMAR coating : 
The lens we used have BMAR coat. 
The residual reflectance of the lens has a wavelength dependence, as shown in Fig.\ref{fig6}. 
The BMAR coating does not work as well at longer wavelengths. 
Ghost images may have appeared 
and the brightness in the dark region may have increased because of the increase in reflected light. 
In principle, this problem can be solved by replacing the lens with mirrors.
Therefore, we are preparing a mirror optics system, which 
requires drastic change to replace the one-dimensional optical bench with the two-dimensional testbed. 
\end{enumerate}
               %
               %
               %
               %
               %
               %
               %


               %
               %
               %
               %
          
\begin{figure}[tbp]
                       \begin{center}
                        \FigureFile(80mm,80mm){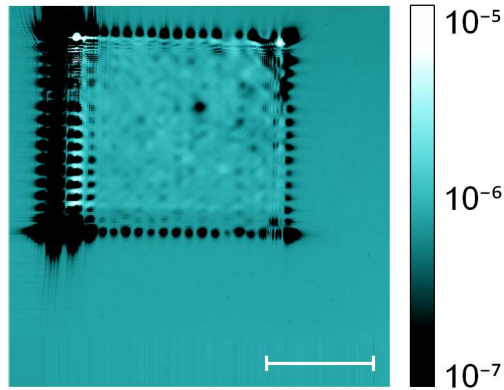}
                       \end{center}
                       \caption{An image of the result of PSF subtraction using two raw coronagraphic images at 650nm and 750nm. 
                       The scale bar is $10\lambda/D$. 
                       }
                       \label{fig7}
                      \end{figure}

         Since we had both multi-color/broadband results, we applied the PSF subtraction at different wavelengths. 
         In principle, PSF subtraction of images at different wavelength cannot completely remove WFE, 
         even static WFE, because WFE has wavelength dependence. 
         On the other hand, if wavelength difference is small, noise pattern is expected to be similar. 
         So we consider PSF subtraction has practical effectiveness for improving contrast (e.g., \cite{Biller}). 
         We used images at 650nm and 750nm which had only a small amount of ghosting. 
         Because the image size is proportional to $\lambda/D$, 
         the image at 650nm was enlarged by 750/650 to make it the same size as the image at 750nm. 
         The positions were aligned by using a bright diffraction pattern outside the dark region as a guide. 
         The intensity was adjusted to the peak of the core image at 750nm. 
         We adjusted the size, position and intensity as described above and subtracted the image at 650nm from image at 750nm. 
         As a result, the areal mean of the dark region had a positive value. 
         In other words, the dark region in 750nm was brighter than the dark region in 650nm. 
         We believe this is because the ghost image is more pronounced at longer wavelengths. 
         The contrast between the peak of the core at 750nm and the areal mean of the dark region after PSF subtraction was 
         7.8$\times 10^{-7}$. 
         The contrast was a little better than 1.1$\times 10^{-6}$ which is the raw contrast of 750nm, 
         though the improvement in contrast was not as much as that obtained in the PSF subtraction experiment with the He-Ne laser. 
         Fig.\ref{fig7} shows no significant elongated pattern expected from difference in $\Delta\lambda/\lambda_c$ at 650nm and 750nm. 
         If we use images at longer wavelengths, the ghosting increases. 
         Thus, there is a need for the installation of a mirror optics system to improve the effects of ghost images.

        
         %
         %
         
         It is beneficial to compare this work with other experiment using a binary-shaped pupil mask coronagraph. 
         \cite{Belikov} achieved the contrast of $4\times 10^{-8}$ using visible laser, 
         and $\sim10^{-7}$ contrast using broadband light source 
         with speckle nulling in a small area from 4$\lambda/D$ to 9$\lambda/D$. 
         On the other hand, our experiments in this paper were performed without speckle nulling. 
         The works shown in this paper resulted  the contrast of $2.3 \times 10^{-7}$ using He-Ne laser, 
         which is just close to the limit of the mask design, and $\sim10^{-6}$ contrast using broadband, 
         in a larger area, from 3$\lambda/D$ to 30$\lambda/D$. 
         Therefore, these works are complementary to each other. 
         It is a possible future work of us to introduce speckle nulling 
         in order to reach higher contrast \citep{Kotani}. 
         In such case, the design of our Mask3 presented in \cite{Enya2008a} can be useful because the contrast 
         produced by this mask is designed to be $10^{-10}$. 
         
         %
         %

 \section{Conclusion}
  We conducted experiments
on a binary-shaped pupil mask coronagraph, which, in principle, should work at all wavelengths. 
The experiments were carried out in a vacuum chamber to improve stability. 
Active wavefront control is not applied in this work. 
As a result of the improved stability, 
we demonstrated an improvement in contrast to 1.3$\times 10^{-9}$ by subtracting the PSF at the He-Ne laser wavelength. 
From the results of multi-color/broadband experiments, 
we verified that significant improvement in contrast 
at various wavelength bands can be produced with the binary-shaped pupil mask coronagraph, 
compared with non-coronagraphic optics. 
The obtained contrasts were close to the contrast between the sun and the earth 
which is $\sim10^{-6}$ in the mid-infrared wavelength region. 
Consequently, this study suggests that a binary-shaped pupil mask coronagraph can be applied to coronagraphic observation 
by SPICA and by other telescopes. 

%
%

\begin{table}
  \caption{Property of light sources and ND filters.}\label{table1}
  \begin{center}
    \begin{tabular}{lccc}
      \hline
Light source & $\lambda_c$  & $\Delta\lambda$ & ND transmissivity\footnotemark[$*$] \\ 
             &  (nm)        & (nm)            & (\%) \\
\hline
He-Ne laser    &  632.8   & 	-      &  	0.0088  \\  
SLED           &  650     & 	 8     &  	0.016  \\ 
SLED           &  750     & 	21     &  	0.25  \\ 
SLED           &  800     &  	25     &  	0.30  \\ 
SLED           &  850     &  	55     &  	0.32  \\ 
\hline
   \multicolumn{4}{@{}l@{}}{\hbox to 0pt{\parbox{85mm}{\footnotesize
       \par\noindent
       \footnotetext{}$*$ Total transmission through the two ND filters at $\lambda_c$.
     }\hss}}
    \end{tabular}
  \end{center}
\end{table}


\bigskip


First of all, we are grateful to the pioneers of the checker-board pupil mask coronagraph, 
especially R. J. Vanderbei. 
This research was partially supported by 
the Japan Aerospace Exploration Agency and the Ministry of Education, 
Science, Sports and Culture, Grant-in-Aid.
The pupil mask was supported by the Nano-Processing Facility of the Advanced Industrial Science and Technology. 
We thank SIGMA KOKI CO., LTD. for providing BMAR data. 
The first author acknowledges support from the JSPS Research Fellowship. 
       

%
%



\end{document}